\documentclass[aps, pra, twocolumn, reprint, nofootinbib, longbibliography]{revtex4-1}

\usepackage{lmodern}
\usepackage{sistyle}
\usepackage{graphicx}
\usepackage{amsmath}
\usepackage{amssymb}
\usepackage{upgreek}
\usepackage{mathrsfs}
\usepackage{verbatim}   
\usepackage{subfigure}  
\usepackage{hyperref}
\usepackage{xcolor}
\hypersetup{
	colorlinks,
	linkcolor={red!95!black},
	citecolor={blue!100!black},
	urlcolor={green!45!black}
}
\usepackage{verbatimbox}

\newcommand{\us}{\upmu\text{s}}

\newcommand{\Fig}[1]{Fig.~\ref{fig#1}}
\newcommand{\Figure}[1]{Figure \ref{fig#1}}

\begin{document}

\title{Simple High-Bandwidth Sideband Locking with Heterodyne Readout}

\author{Christoph Reinhardt}
\author{Tina M\"uller}
\author{Jack C. Sankey}
\affiliation{Department of Physics, McGill University, Montr\'{e}al, Qu\'{e}bec, H3A 2T8, Canada}

\email{jack.sankey@mcgill.ca}


\begin{abstract}
We present a robust sideband laser locking technique that is ideally suited for applications requiring low probe power and heterodyne readout. By feeding back to a high-bandwidth voltage controlled oscillator, we lock a first-order phase-modulation sideband to a table-top high-finesse Fabry-Perot cavity, achieving a feedback bandwidth of \SI{3.5}{MHz} with a single integrator, limited fundamentally by the signal delay. The directly measured transfer function of the closed feedback loop agrees with a model assuming ideal system components, and from this we suggest a modified design that should realistically achieve a bandwidth exceeding \SI{6}{MHz} with a near-causally limited feedback gain of $4\times 10^7$ at $\SI{1}{kHz}$. The off-resonance optical carrier is used for alignment-free heterodyne readout, alleviating the need for a second laser or additional optical modulators.   
\end{abstract}

\maketitle

%
\section{Introduction}
\noindent
A common goal in precision optics is to use feedback \cite{Bechhoefer2005Feedback} to stabilize (lock) the frequency of a laser to that of an external system such as a Fabry-Perot resonator \cite{Drever1983Laser} or atomic transition \cite{Dschao1980I2, Cerez1980He}. This can be used either to stabilize the laser itself or to monitor the dynamics of the external system. In all feedback schemes, it is desirable to achieve the largest possible closed-loop gain (i.e.~the degree to which noise can be suppressed) and a dynamic range (headroom) sufficient to compensate for the largest fluctuations. 

Due to causality, the gain of a feedback loop is ultimately limited by the speed with which corrections can be applied (i.e.~the loop's bandwidth), which in turn is fundamentally limited by the delay of the signal propagating through the loop \cite{Bechhoefer2005Feedback}. In many situations, the achievable gain is practically limited by other nonidealities. For example, one means of tuning laser frequency is to mechanically stretch an optical path, but the bandwidth is then practically limited by the structure's mechanical resonances. For this reason, typical low-noise, mechanically tuned lasers (e.g.~commercial Nd:YAG) achieve control bandwidths limited to $\sim$\SI{100}{kHz}. Faster feedback can be achieved via electronic control of the laser's pump. Diode lasers, for example, readily achieve $\sim$5 MHz bandwidth using pump feedback to stabilize to an external cavity \cite{Schoof2001Reducing}, and as such this technique is routinely used as a first stage in reducing their comparatively large noise; the combined system, however, is then subject to the mechanical bandwidth of the external cavity. Cavity length stabilization has improved in recent years, achieving \SI{180}{kHz} using short-travel piezo actuation \cite{Briles2010Simple} and now up to $\sim$\SI{700}{kHz} with the incorporation of photothermal tuning \cite{Brachmann2016Photothermal}. 

Instead of controlling the frequency of the light generated by a laser, one can also shift the frequency after emission. For visible wavelengths, this is often accomplished with an acousto-optical modulator (AOM), which can achieve $\sim$\SI{200}{kHz} feedback bandwidth \cite{Kessler2012A} and $\sim$MHz-scale headroom. At near-infrared (telecom) wavelengths, low-cost fiber modulators are more commonly employed. Using serrodyne techniques, wherein a voltage-controlled oscillator (VCO), nonlinear transmission line (NLTL), and electro-optical modulator (EOM) generate a saw-tooth phase that effectively shifts the carrier frequency \cite{Houtz2009Wideband, Kohlhaas2012Robust}, or single-sideband modulation (SSM), wherein a Mach-Zehnder interferometer shifts a small portion of the carrier \cite{Gatti2015Wide}, it is routine to achieve several-MHz feedback bandwidth and well over \SI{100}{MHz} headroom.

A second common goal in precision optics is to perform heterodyne readout \cite{Protopopov2009Laser}, wherein a weak ``probe'' beam is overlapped with a strong optical local oscillator (LO) detuned by an electronically-measurable frequency. When measured by a photodiode, the beating between these beams produces an amplified electronic signal from the probe with a spectrum shifted to the LO detuning, thereby providing access to the signal's amplitude and phase quadratures. 

Here we present and characterize a simple, low-cost, high-bandwidth, post-emission laser locking technique with built-in heterodyne readout. In complement to serrodyne and SSM systems, this approach relies on a high-speed VCO and optical modulator to control the frequency, and can be implemented with any laser. In contrast to serrodyne systems, this does not require a precise saw-tooth or high-bandwidth EOM, and, similar to SSM, shifts only a fraction of the laser light. In contrast to both, the carrier is exploited as an optical LO for heterodyne readout, and since this follows the same optical path, no alignment or relative path stabilization is required. Using the test ports of our chosen electronics, we directly measure the frequency-dependence of the closed-loop gain, demonstrating a delay-limited feedback bandwidth of \SI{3.5}{MHz} and headroom exceeding \SI{500}{MHz} ($\sim$\SI{1}{GHz} should be possible with this VCO/EOM combination, at the expense of added amplitude noise). The measured gain matches a simple model based on ideal components, and from this we propose a modified setup that should realistically achieve a gain of $4\times 10^7$ at \SI{1}{kHz} (\SI{6.6}{MHz} bandwidth). Section \ref{sec:review} briefly reviews requisite concepts in laser feedback. Section \ref{sec:PDH} then introduces the ``Pound-Drever-Hall'' method for generating an error signal \cite{Pound1946Electronic,Drever1983Laser,Black2001An} (including a derivation of its dynamical response), along with a simple electronic modification enabling heterodyne readout. We then present the technical details of our ``proof-of-concept'' system in Sec.~\ref{sec:technique}, characterize its closed-loop performance in Sec.~\ref{sec:performance}, and conclude in Sec.~\ref{sec:discussion}.

\section{A Brief Review of Laser Feedback}\label{sec:review}

\noindent
All frequency stabilization schemes rely on (i) the generation of an ``error'' signal proportional to the detuning $\delta$ between the laser and the external system, and (ii) the processing and routing of this signal to a port capable of adjusting $\delta$ to compensate \cite{Bechhoefer2005Feedback}. \Figure{1}(a) shows a generic diagram of a ``typical'' feedback loop for locking a laser to a cavity resonance. Environmental noise (e.g.~vibrations, laser noise) introduces a nominal detuning $\delta_n$, which is subsequently converted to an optical signal by the cavity ``$C$'', translated into an electrical signal by a photodiode ``$D$'', and modified by assorted electronic components and amplifiers ``$-A$'', before being sent to a ``feedback'' port ``$F$'' to compensate. This correction is added to the environmental noise, resulting in a relationship for the \emph{actual} detuning $\delta = \delta_n-CDAF\delta$, where $C$, $D$, $-A$, and $F$ are the complex, frequency-dependent complex gains (transfer functions) of the cavity, diode, electronics, and feedback port. Solving for $\delta$ yields
\begin{equation}\label{eq:feedback}
\delta = \frac{\delta_n}{1+CDAF}.
\end{equation}
This immediately highlights the central concerns for stabilization. First, the ``closed-loop gain'' $G\equiv CDAF$ should be made as large as possible to cancel the environmental noise. For $|G| \gg 1$, the overall phase $\phi_G$ does not matter, but if $G$ approaches $-1$ at a some frequency, then the noise at that frequency is \emph{amplified}. This places unavoidable limits on $G$ for the following reasons: (i) any delay $t_d$ in the signal path multiplies $G$ by the phase factor $e^{-i\omega t_d}$, forcing $\phi_G=-\pi$ at finite frequencies, regardless of what electronics are chosen for $A$, (ii) stability concerns impose that the magnitude of the gain at the lowest of these frequencies $\omega_{-\pi}$ should be less than 1, and (iii) causality places an upper bound $|G| < \omega_{-\pi}^2/\omega^2$ on how much the gain can increase below this point \cite{Bechhoefer2005Feedback}. Since most noise occurs at low frequencies, it is therefore desirable to make $\omega_{-\pi}$ large, and to engineer a feedback circuit such that $G$ increases as rapidly as possible below that frequency.

\begin{figure}[!ht]
	\centering
	\includegraphics[width=0.98\columnwidth]{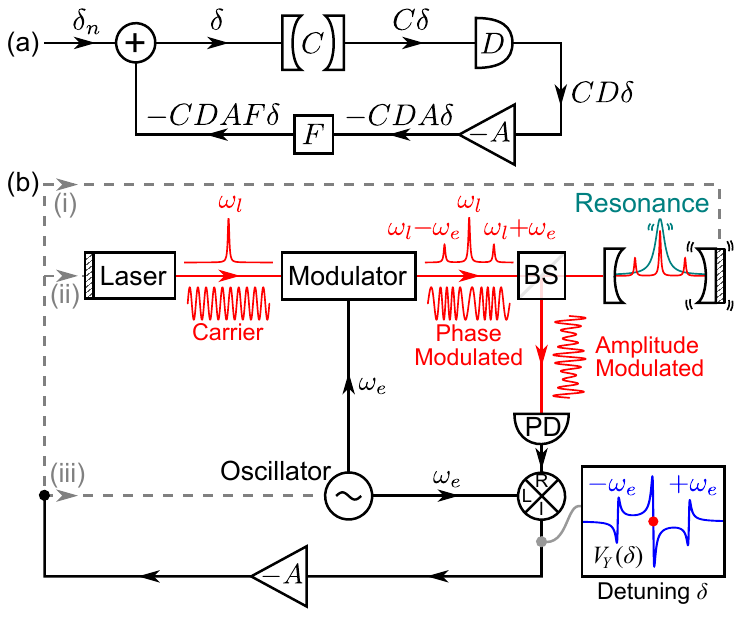}
	\caption{Feedback stabilization. (a) Generic control loop for stabilizing a laser's detuning $\delta$ from the resonance of an optical cavity. Noise $\delta_n$ enters, is converted to an optical signal by the cavity (transfer function $C$), collected by a diode ($D$), manipulated by electronics ($-A$) and sent to a ``feedback'' port ($F$). (b) Practical implementation using Pound-Drever-Hall readout. Straight red lines represent optical paths, straight black lines represent electrical paths, and dashed gray lines show potential feedback paths. The laser is phase-modulated, lands on a beam splitter (BS) and interacts with the cavity, which converts phase to amplitude modulation. This is recorded with a photodiode and mixed (demodulated) with a local oscillator. Inset shows the resulting steady-state voltage $V_Y(\delta)$, with a red dot indicating a stable lock point. The manipulated signal can be fed back to (i) the cavity length or (ii) the laser frequency. Feeding back to (iii) the oscillator frequency only adjusts the sidebands.}
	\label{fig1}
\end{figure}

A readout of $\delta$ (the error signal) can be obtained by several methods. A high-finesse optical cavity having length $L$, input mirror (power) transmission $T$ and reflection $R$, and power ringdown time $\tau$ has an overall field reflection coefficient (see Supplementary Material)
\begin{equation}\label{eq:cavity}
r(\delta)\approx \frac{c\tau T/L}{1+i2\tau\delta}-\sqrt{R}.
\end{equation}
The reflected power ($\propto$$|r|^2$) therefore follows a Lorentzian line shape, and on resonance ($\delta$=$0$), cannot on its own be used for feedback, since it does not provide information about the sign of $\delta$. One can of course generate a bipolar error signal by tuning the laser away from resonance \cite{Barger1973Frequency}, but this technique couples laser power fluctuations to detuning errors. However, the phase of $r(\delta)$ does vary linearly on resonance, and has been extracted via phase modulation \cite{Drever1983Laser}, heterodyne \cite{Protopopov2009Laser,Danilishin2012Quantum}, and homodyne \cite{Heurs2010Homodyne} schemes, wherein the mode of interest interferes with one or more reference beams having different frequency or phase. Other techniques employ a second cavity mode as a reference, for example a mode of different polarization \cite{Hansch1980Laser} or a higher order spatial mode \cite{Wieman1982Laser, Shaddock1999Frequency}. The ubiquitous and powerful ``Pound-Drever-Hall'' technique \cite{Drever1983Laser} is discussed in the following section.

\section{Modified Pound-Drever-Hall Readout and Dynamical Response}\label{sec:PDH}
\noindent
A common method for on-resonance laser stabilization is the ``Pound-Drever-Hall'' (PDH) technique \cite{Pound1946Electronic,Drever1983Laser}, a diagram of which is drawn in \Figure{1}(b). Stated briefly, this technique effectively amounts to dithering the laser frequency with an electro-optical modulator (EOM) and measuring the induced modulation in the reflected power to infer the \emph{slope} of $|r(\delta)|^2$ (or Im[$r$]) \cite{Black2001An}. The resulting error signal (inset blue curve, near red dot) can then be manipulated with electronics ($-A$) and fed back to either (i) the cavity length or (ii) the laser frequency, as described above. Feeding back to a voltage-controlled oscillator (VCO, iii) will not adjust the \emph{carrier} frequency (or $\delta$) in this configuration, but can be used to lock a \emph{sideband} to the cavity as discussed in Sec.~\ref{sec:technique}. An elegant, pedagogical derivation of the steady-state error signal ($V_Y$ in \Fig{1}(b)) from this system can be found in Ref.~\cite{Black2001An}. This accurately captures the system's ability to convert low-frequency detuning noise into an error signal, but breaks down when the detuning $\delta(t)$ contains frequencies comparable to the cavity's linewidth $1/\tau$. A straightforward means of deriving the \emph{dynamic} response \cite{Rakhmanov2002Dynamic} for small deviations $\delta$ about the lock point is to propagate a laser ``noise'' component through an EOM, cavity, diode, and demodulation (mixer) circuit in Fig.~\ref{fig1}(b) to extract a combined transfer function, as follows. 

Suppose there exists a frequency noise component at frequency $\omega$ that is the real part of $\Omega(t)=\Omega_n e^{i\omega t},$ where $\Omega_n$ is a constant. This corresponds to phase modulation $\phi(t)=\phi_n \sin(\omega t)$, where $\phi_n = \Omega_n/\omega$. If this light is fed through a phase modulator (EOM) driven by voltage $V_\text{osc} = V_e \sin(\omega_e t)$, the field landing on the cavity is then
\begin{eqnarray}\label{eq:field}
E = E_l \cos\left(\omega_l t + \phi_e \sin\omega_e t + \phi_n \sin\omega t\right)
\end{eqnarray}
where $E_l$ is a constant amplitude and $\phi_e\propto V_e$ according to the efficiency of the EOM. Assuming all modulations are small ($\phi_e,\phi_n\ll 1$), Eq.~\ref{eq:field} can be written as the sum of a ``carrier'' at frequency $\omega_l$, 4 first-order sidebands ($\omega_l\pm \omega_e$ and $\omega_l\pm\omega$) and 8 second-order sidebands ($\omega_l\pm 2\omega_e$, $\omega_l\pm 2\omega$, $\omega_l \pm \omega_e \pm \omega$, and $\omega_l \pm \omega_e \mp \omega$). If we also assume the modulator frequency is large compared to the cavity linewidth and noise frequency ($\omega_e \gg 1/\tau,\omega$), and the \emph{carrier} is on resonance, only five beams ($\omega_l$, $\omega_l \pm \omega$, and $\omega_l \pm 2\omega$) acquire a significant change in magnitude and phase upon reflection, as per the cavity response (Eq.~\ref{eq:cavity}). When the 13 reflected beams land on a photodiode, they produce a time-averaged photocurrent $\propto$$\langle E^2\rangle$ containing all frequencies ($\ll \omega_l$) within the photodiode's bandwidth. If this signal is then mixed with the original oscillator voltage $V_\text{osc}$, the mixer output is proportional to $\langle E^2\rangle \sin(\omega_e t)$, and an appropriately chosen low-pass filter can then eliminate all terms except those having frequency near $\omega$. After some bookkeeping (see Supplementary Material), the transfer function for converting a frequency noise $\Omega$ to an error signal $V_Y$ is
\begin{equation}\label{eq:dynamic-PDH}
\frac{V_Y}{\Omega} \approx 
-\frac{2\phi_e E_l^2\beta \tau^2}{1+2i\tau\omega} 
\end{equation}
where constant $\beta$ includes a combination of cavity constants and the conversion efficiencies of the diode and mixer.\footnote{Note the diode and mixers employed below have large bandwidths, and are assumed to have frequency-independent efficiencies for simplicity. This assumption is validated for our chosen components, as discussed below.} The interpretation of this result is straightforward. Assuming $\phi_n\ll 1$ restricts the $V_Y(\delta)$ to the region of linear response (i.e.~near the red dot in Fig.~\ref{fig1}(b)). The resulting transfer function sensibly scales with the laser power and dither amplitude \cite{Drever1983Laser,Black2001An}, and the cavity's amplitude ringdown time $2\tau$ imposes a low-pass filter on the readout \cite{Rakhmanov2002Dynamic}.

Equation \ref{eq:dynamic-PDH} also motivates the use of a ``proportional-integral'' (PI) amplifier for feedback electronics. A PI amplifier has a transfer function
\begin{equation}
\label{eq:PI-amp}
A_{PI} = G_0\frac{1+i\omega/\omega_{PI}}{1/g+i\omega/\omega_{PI}}
\end{equation}
where $G_0$ is an overall scaling factor, $\omega_{PI}$ is a ``PI corner'' frequency, above which the response changes from integrator-like to proportional, and $g$ is a gain limit at low frequencies. Often (especially while locked) the gain limit is removed ($1/g\rightarrow0$), in which case $A_{PI}\rightarrow G_{0}\left(1-i\frac{\omega_{PI}}{\omega}\right)$; when combined with the readout transfer function (Eq.~\ref{eq:dynamic-PDH}), the choice $\omega_{PI}=1/2\tau$ then results in a partial-loop transfer function 
\begin{equation}\label{eq:nearly-closed-gain}
\frac{V_Y}{\Omega}A_{PI} = \frac{\phi_e E_l^2 \beta \tau^2 G_0}{i \tau\omega}
\end{equation}
The total system behaves like an integrator over all frequencies, with increasing gain at low frequencies. The overall phase is also far from $-\pi$, preventing the system's delay factor $e^{-i\omega t_d}$ from forcing the closed-loop gain below 1 at a low frequency. This also provides ``wiggle room'' for loop nonidealities such as indirectly driven resonances that can cause a temporary excursion in phase (see, e.g.,~Ref.~\cite{Briles2010Simple}). However, even if the bandwidth of the feedback port $F$ is effectively infinite and / or we have precisely compensated for all of its artifacts, the ultimate gain is limited by the signal delay $t_d$ -- in this case from the output of the EOM to the cavity, back to the diode, through the electronics, and through the feedback port -- which forces the closed-loop gain to be less than 1 at frequency $\omega_{-\pi} < \pi/4t_d$ for this choice of electronics.

As mentioned (and discussed below) it is also possible to lock the first-order sidebands ($\omega\pm\omega_e$) to the cavity. Following the same analysis for the case of either sideband resonant with the cavity produces a transfer function (see Supplementary Material)
\begin{equation}
\label{eq:dynamic-sideband}
\frac{V_{Y,\pm}}{\Omega} \approx 
\frac{\phi_e E_l^2 \beta \tau^2}{1+2i\tau\omega} 
\end{equation} 
which is inverted and half as large as the carrier-resonant case (Eq.~\ref{eq:dynamic-PDH}), consistent with the slope of the steady state solution ($V_Y$ in \Fig{1}(b)) at $\delta=\pm\omega_e$ \cite{Drever1983Laser, Black2001An}. 

\begin{figure*}[htb]
	\includegraphics[width=0.95\textwidth]{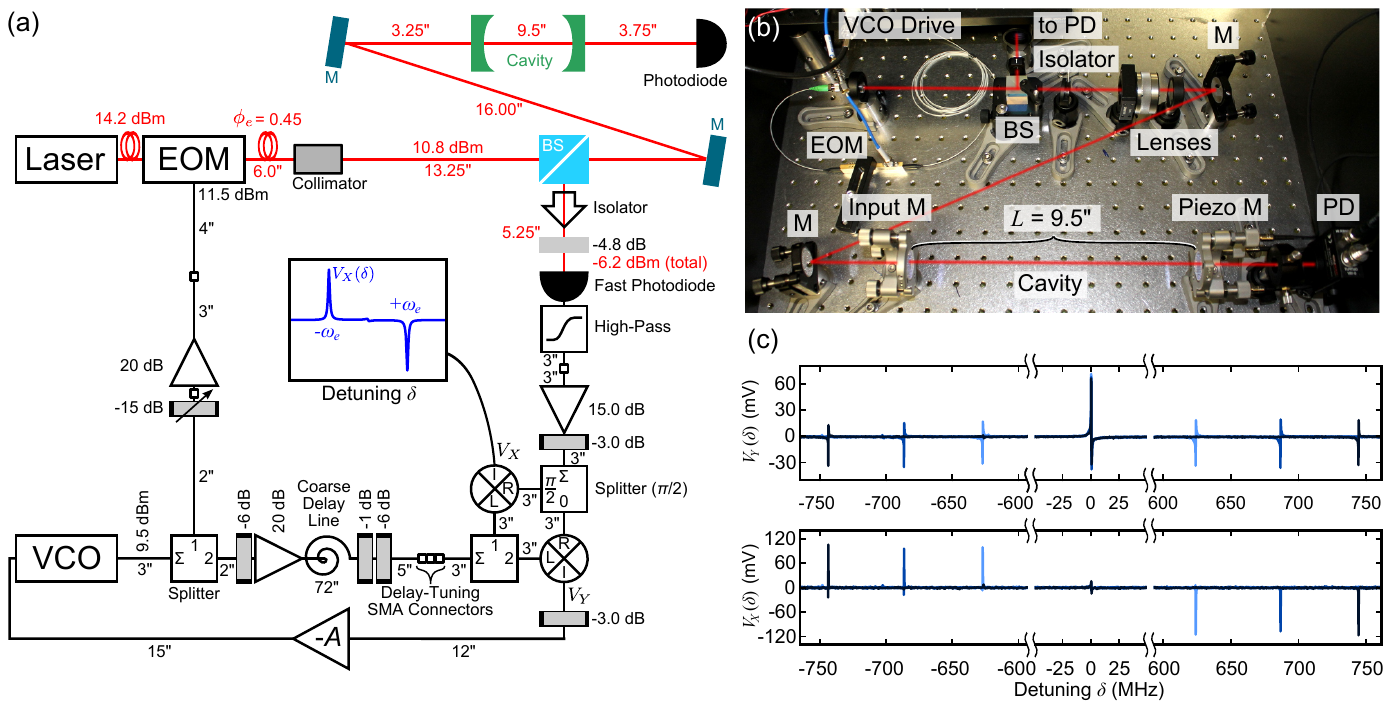}
	\caption{Sideband locking with heterodyne readout. Many parts from Thorlabs (TL) and Minicircuits (MC). (a) A VCO (MC ZX95-1600W-S+) signal is split (MC ZX-10-2-20-S+) and amplified (MC ZX60-4016E-S+) feeding both an EOM (TL LN65-10-P-A-A-BNL-Kr with shortened output fiber) and, after delay, the electronic LO (``L'') ports of two mixers (MC ZFM-5X-S) for quadrature readout. A fiber laser (Koheras Adjustik E15) feeds a 14.2-dBm (26.3 mW) carrier through the EOM, producing a \SI{10.8}{dBm} carrier and \SI{-2.2}{dBm} (5\%) sidebands, tuned by a variable attenuator (MC ZX73-2500-S+, $\sim$\SI{15}{dB}) leading to the EOM. Once collimated (TL F260APC-1550), the beam passes through a beam splitter (BS, TL BS018 50:50), mode-matching lenses (\SI{-5}{cm} and \SI{10}{cm} focal length, shown in (b)), and steering mirrors (M) before landing on a cavity comprising a flat (Newport 10CM00SR.70F) and curved (Newport 10CV00SR.70F) supermirror, the second of whose position is swept by a piezo mirror mount (TL K1PZ). The transmitted beam is focused on a photodiode (PD, TL PDA10CF), while the reflected beam is rerouted by the BS, passes through a free-space isolator (TL IO-2.5-1550-VLP) to eliminate standing waves, and is focused upon a 2-GHz-bandwidth low-noise photodiode (PD, Femto HSA-X-S-2G-IN). The signal's low-frequency noise ($<20$ MHz) is eliminated with a high-pass (MC SHP-20+), before amplification (MC ZX60-P105LN+) and splitting by a $\pi/2$ splitter (MC ZX10Q-2-13-S+). The phase-shifted signals are fed to the mixer's RF (``R'') ports for demodulation to the IF (``I'') ports. The ``phase'' quadrature ($V_Y$) is fed into a PI amplifier ($-A$, New Focus LB1005) for feedback to the VCO. Inset shows the predicted steady-state voltage of the amplitude quadrature $V_X(\delta)$. (b) Photograph of optical setup. (c) Simultaneously acquired phase and amplitude quadratures from points (i) and (ii) in (a), respectively, for three different VCO control voltages: 0 V (lightest), 0.8 V, and 1.8 V (darkest). The cavity length is swept quickly to avoid run-to-run variations due to vibrations, and to show transient signals common to high-finesse cavities (see text).}
	\label{fig2}
\end{figure*}

Finally, similarly propagating an \emph{amplitude} noise component through this system (i.e., setting $E_l\rightarrow \left(1+\operatorname{Re}[\epsilon]\right) E_l$, where $\epsilon(t) = \epsilon_n e^{i\omega t}$ with constant $\epsilon_n\ll 1$) has no impact on $V_Y$ or $V_{Y,\pm}$. However, introducing a relative $\pi/2$ phase shift between the mixer's LO and signal ports provides a measurement of the other quadrature $V_X$, which carries amplitude information. When locked to either sideband, the amplitude quadrature transfer function is (see Supplementary Material)
\begin{equation}
\label{eq:dynamic-sideband-amp}
\frac{V_{X,\pm}}{\epsilon} \approx \mp \phi_e E_l^2 \beta \tau
\frac{1+i\tau\omega}{1+2i\tau\omega}
\end{equation}
for the upper or lower sidebands resonant with the cavity, respectively. We note that, in contrast to $V_{Y,\pm}$, the amplitude quadrature $V_{X,\pm}$ is influenced by the off-resonance sideband. Hence, adding a second phase-shifted mixer (or using an IQ mixer) enables heterodyne readout with no additional lasers, optical modulators, or alignment. Conveniently, the steady-state form of this quadrature, discussed below and shown in Fig.~\ref{fig2}, also provides a simple means of verifying which sideband is locked to the cavity (along with an independent estimate of how well it is locked).

\section{Apparatus for Sideband Locking with Heterodyne Readout}\label{sec:technique}

\noindent
\Figure{2}(a) shows our test setup for locking a first order sideband (at $\omega_l\pm\omega_e$) to a cavity resonance. Sidebands are created with a fiber EOM driven at $\omega_e$ by a VCO with \SI{90}{MHz} modulation bandwidth and 0.65-\SI{1.75}{GHz} tuning range. Light from the EOM passes through a beam splitter (BS) and mode-matching optics (shown in (b)), reflects from the cavity, and is collected by a high-bandwidth photodiode. The resulting signal is filtered and amplified before passing through a power splitter that produces a phase shift of 0 and $\pi/2$ at its outputs. These two signals are separately mixed with that of the VCO to produce $V_X$ and $V_Y$. The VCO output is split prior to the EOM, delayed, and used as the electronic LO for both mixers. In order to maintain a fixed phase between the mixers' LO and signal ports over the full range of VCO frequencies $\omega_e$, the delay between the two signal paths must match. Any difference $\Delta t_d$ produces a relative phase $\omega_e \Delta t_d$ that must remain small compared to $\pi/2$ at the highest VCO frequency. Here this imposes that $\Delta t_d \ll \pi/2 \omega_e \sim 1$ ns, corresponding to a free-space path difference $\ll 30$ cm; this is mostly compensated for with a combination of cables and extension adapters (\Fig{2}(a)), with mm-scale fine tuning of the photodiode's position. The higher precision required for larger-$\omega_e$ systems can be easily implemented with the diode optics mounted on a translation stage.

\Figure{2}(b) shows a photograph of the optical path; the electronics are mounted on a nearby platform. The detuning $\delta$ between the laser and cavity can be widely adjusted with long-travel piezos in the second mirror mount (``Piezo M''). \Figure{2}(c) shows a diagnostic measurement of $V_Y(\delta)$ and $V_X(\delta)$ recorded during cavity length sweeps for a few values of $\omega_e$. Each sweep was performed ``quickly'' (16 ms over the full range) to reduce run-to-run variations from the ambient vibrations of the test cavity. The insensitivity of the quadrature readout to $\omega_e$ indicates the delay is matched (see Supplementary Material for a larger range). The cavity has a power ringdown time $\tau=1.2\pm0.1~\us$ (finesse 4700$\pm$400), and so these fast sweeps produce a transient response \cite{Lawrence1999Dynamic} resulting in a measured $V_Y$ (top plot of (c)) that is consistently not symmetric about $V_Y=0$, and a measured $V_X$ (bottom plot of (c)) that deviates from a simple peak. This artifact can be highly misleading when tuning the relative delay, and so rather than trying to symmetrize $V_Y$, we recommend slowly modulating $\omega_e$ while quickly sweeping the cavity, and adjusting $\Delta t_d$ to produce a signal shape that does not vary with $\omega_e$.

The error signal $V_Y$ is then fed through a tunable PI amplifier having the transfer function of Eq.~\ref{eq:PI-amp}, with $\omega_{PI} = 110$ kHz and $g = 105 =\SI{40}{dB}$ (measured) before finally being fed back to the VCO. Due to the sidebands' opposed frequency response, one is always stabilized by this feedback and one is always destabilized; here we (arbitrarily) lock the upper sideband (verified by the negative value of $V_X$). Despite the open-air design and flagrant disregard for vibration isolation, this system readily locks and remains so indefinitely.\footnote{The lock is impervious to chair scoots, door slams, claps, and shrieks, but fails if the table surface is tapped with a wrench.}

\section{Performance}\label{sec:performance}
\noindent
Once locked, we increase the feedback gain $G_0$ until the system rings (at $\sim$\SI{3}{MHz} for this implementation), indicating that the gain at $\omega_{-\pi}$ has exceeded unity, with $\omega_{-\pi}\sim$ \SI{3}{MHz}. We then reduce $G_0$ until the remaining noise in $V_Y$ is minimized. The most sensitive estimate of $V_Y$ is achieved by referring the PI amplifier's output back to its input using its known (measured) transfer function; together with an independent measurement of the error signal slope on resonance $2\pi\times \partial_\delta V_Y = 388\pm40$ mV/MHz, we estimate that the stabilized RMS detuning noise $\delta_\text{RMS}/2\pi$ is below \SI{70}{Hz} (0.0005 cavity resonance full-widths). This is a factor of 3000 lower than the pre-stabilized value of $240$ kHz (1.6 fullwidths, corresponding to \SI{0.3}{nm} RMS cavity length noise), as estimated directly from the PI output and the VCO specifications (\SI{52}{MHz/V}). \Figure{3}(a) shows the power spectral densities of these two inferred detuning signals. The square root of their ratio provides a basic estimate of the closed-loop gain magnitude $|G(\omega)| \sim 1000$ at $\omega/2\pi=\SI{1}{kHz}$.

To directly measure $G(\omega)$, we inject a small amount of ``noise'' into the locked system and observe how it is suppressed. The PI amplifier provides a second (inverted) input, and an isolated input monitor for independently measuring the in-loop error signal. Using a lock-in amplifier, we apply an oscillatory signal $V_n$ of frequency $\omega$ to this input and record both quadratures of the error signal $V_Y$ at $\omega$ (correcting for the transfer functions between the input and error monitor, as well as the lock-in and its measurement cables). Using the same analysis of \Fig{1}(a) with $CDAF \rightarrow G$, $\delta_n\rightarrow V_n$, and $\delta\rightarrow V_Y$, we solve for the closed-loop gain $G=V_n/V_Y - 1$, which is plotted in \Fig{3}(b) (blue). Importantly, the observed gain smoothly decreases with $\omega$ (approximately as $1/\omega$), and the phase crosses $-\pi$ at $\omega/2\pi=\SI{3.5}{MHz}$, where $|G|<1$, consistent with the observed ringing frequency. The measurement noise increases at low frequencies due to the reduced signal at high gain. It is worth noting that, despite the addition of sidebands to the VCO output (at $\omega_e\pm \omega$), the measured transfer function of the EOM, cavity, diode, and mixer is identical to that of simple laser frequency noise (see Supplementary Material). 

\begin{figure}[!ht]
	\centering
	\includegraphics[width=0.8\columnwidth]{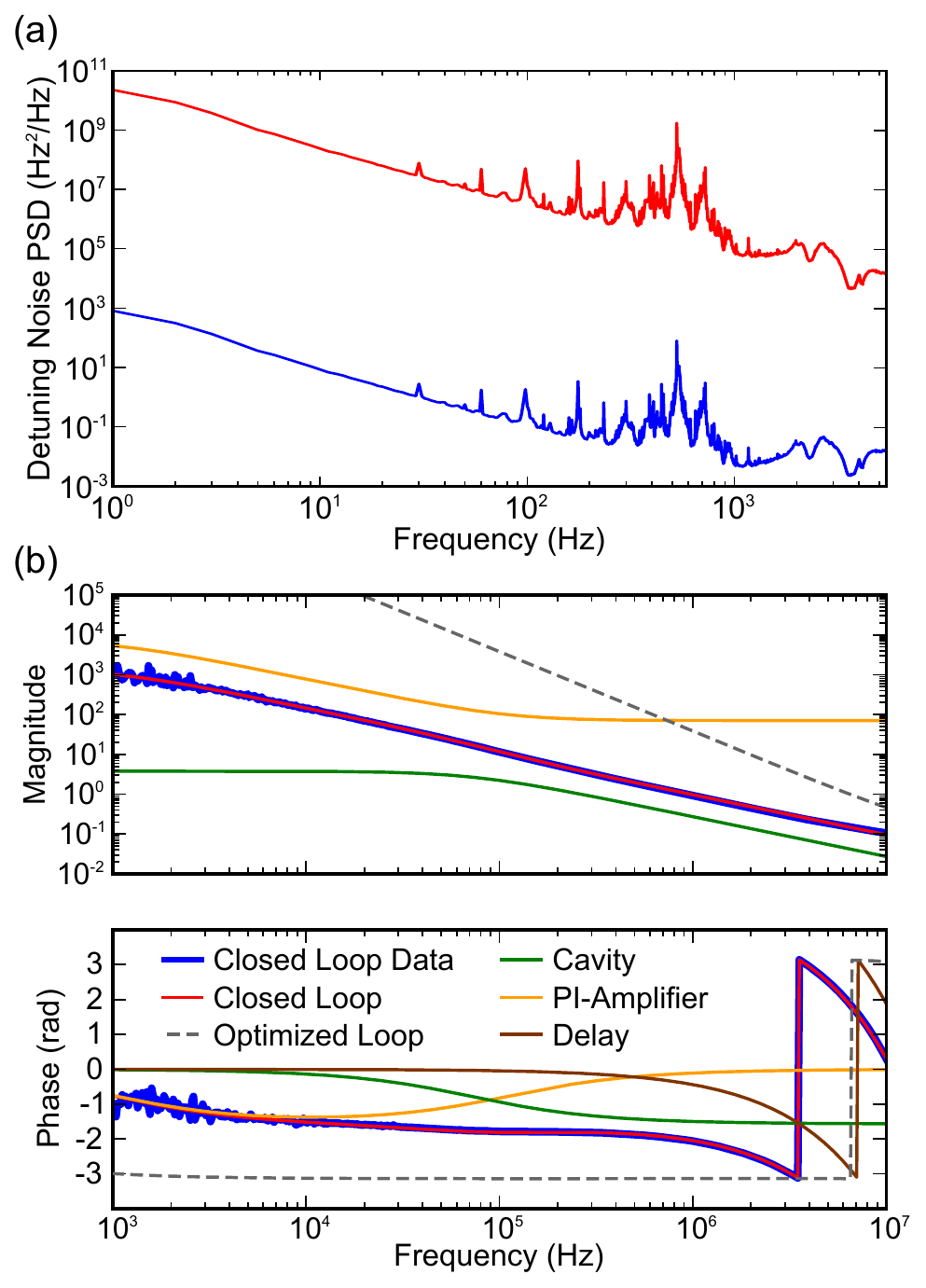}
	\caption{(a) Detuning noise power spectral density (PSD) before and after lock, recorded while locked. The pre-feedback noise (red) is inferred from the proportional-integral (PI) amplifier output and the VCO conversion factor 52 MHz/V, while the post-feedback noise (blue) is inferred from the PI output referred back to its input and the independently measured slope of the error function ($388\pm40$ mV/MHz) at the lock point. (b) Measured (blue) and modeled (red) closed-loop transfer function. The model includes the cavity (green, ring-down time $\tau=\SI{1.1}{\mu s}$), PI amplifier (yellow, $\omega_{PI}$=\SI{110}{kHz}, and g=105), and a delay (brown, \SI{70}{ns}, \SI{52}{ns} from the PI amplifier). Transfer functions of other components are assumed to be ``flat'' on this scale. The gray dashed line shows a closed-loop gain that could be achieved with optimizations: replacement of the PI amplifier and further delay reductions to \SI{10}{ns} and two PI filters, one with $\omega_{PI}/2\pi=\SI{70}{kHz}$, $1/g=0$), and the other with $\omega_{PI}/2\pi=\SI{15}{MHz}$ and $g=10^5$. }
	\label{fig3}
\end{figure}

The red line in Fig.~\ref{fig3}(b-c) represents a simple model for $G(\omega)$ comprising the product of (i) the PI transfer function (Eq.~\ref{eq:PI-amp}) with measured $\omega_{PI}$=\SI{110}{kHz} and $g$=105, (ii) the cavity transfer function (Eq.~\ref{eq:dynamic-sideband}) with $\tau$=\SI{1.1}{\us} (i.e.~one standard deviation below the measured value), (iii) a closed-loop delay $t_d=70$ ns, and (iv) an overall scaling factor chosen to match the measured $G(\omega)$. The yellow and blue curves show the modeled PI and cavity transfer functions alone for reference, and the brown curve shows the phase contribution from the delay. The employed value of $t_d$ is consistent with the signal travel time of the loop, independently estimated to be approximately \SI{68}{ns} from the signal path of the lower VCO loop in Fig.~\ref{fig2}(a): a combined cable and component length of 127" traversed at $2/3$ the speed of light (\SI{16}{ns}) plus the measured internal delay of the PI amplifier (\SI{52}{ns}). The agreement between the model and measurement suggests that the chosen components exhibit no important nonidealities up to $\sim$\SI{10}{MHz}, and that the other components (the EOM, optics, diode, filters, mixers, amplifiers, attenuators, splitters, and connectors) can be assumed to have a flat response, adding a combined delay on the order of nanoseconds at most. 

The phase plot of \Fig{3}(b) highlights that the achieved bandwidth is limited primarily by the delay. Without it, the phase would remain above $-\pi/2$ to a significantly higher frequency, allowing for larger $G_0$. The PI amplifier accounts for 75\% of the delay, implying the greatest gains can be made by replacing it with a faster (albeit less flexible) integrated circuit. Modern amplifiers routinely achieve sub-nanosecond delays, and the requisite PI filters can be realized with passives (capacitors and resistors). It is also straightforward to reduce the optical and electronic lengths: using compact mode-matching optics and shorter cables alone can reduce the delay to $\sim$\SI{10}{ns}. Furthermore, replacing the existing PI filter with two -- one having $\omega_{PI}/2\pi = 70$ kHz and $1/g\rightarrow 0$ and the other having $\omega_{PI}/2\pi=15$ MHz and $g=10^5$ -- for example, would produce a bandwidth of \SI{6.6}{MHz} and (more importantly) a near-causality limited gain $|G(2\pi\times\SI{1}{kHz})|\sim 4\times 10^7$ (Fig.~\ref{fig3}(b), dashed line). This optimization will be the subject of future work.

To estimate the headroom, we change the cavity length $L$ while locked and monitor the output voltage of the PI amplifier; the system remains locked over the full $\sim$\SI{100}{MHz} tuning range presented in Fig.~\ref{fig2}(b), in this case limited by the cavity's small free spectral range: the lower sideband of an adjacent mode eventually becomes degenerate with the locked sideband, spoiling the error signal. Performing the same test on a 5-cm cavity, we find a headroom of \SI{550}{MHz}, limited instead by the maximum output voltage of the PI amplifier (10 V), which covers only half the tuning range of the VCO. A headroom exceeding \SI{1}{GHz} is in principle possible with these components, however, while more headroom is certainly useful for tracking large fluctuations, the frequency-dependencies of the VCO output, EOM, and other electronics will eventually couple these fluctuations to the amplitude of the probe and optical LO beams (see Supplementary Material). Further engineering effort is therefore best spent reducing the system's inherent noise.

\section{Summary}\label{sec:discussion}

\noindent
We have demonstrated a simple technique for locking a first order laser sideband to an optical cavity with a delay-limited feedback bandwidth of \SI{3.5}{MHz} with a single integrator, and a headroom exceeding \SI{500}{MHz}. We directly measured the closed-loop gain, finding excellent agreement with a model based on ideal components, and suggest simple modifications for realizing a gain exceeding $10^7$ at \SI{1}{kHz}. Finally, we note that, by implementing an appropriately weighted sum of $V_X$ and $V_Y$ (or otherwise shifting the relative phase of the mixers' electronic LO and signal ports), it should be possible to create an amplitude-insensitive locking point (i.e.~a zero crossing in the resulting error signal) at arbitrary detuning.

\section{Acknowledgments}
\noindent
We thank Erika Janitz, Maximilian Ruf, Alexandre Bourassa, Simon Bernard, Abeer Barasheed, and Vincent Dumont for helpful discussions. T.M. acknowledges support by a Swiss National Foundation Early Postdoc Mobility Fellowship. The authors also acknowledge computational support from Calcul Qu\'{e}bec and financial support from NSERC, FRQNT, the Alfred P. Sloan Foundation, CFI, INTRIQ, RQMP, CMC Microsystems, and the Centre for the Physics of Materials at McGill.

\bibliography{Alles}  

\end{document}